%% file: sample-base.tex
\documentclass[sigconf]{acmart}

\usepackage{xspace}
\usepackage{hyperref}
\usepackage{xcolor}
\usepackage{adjustbox}
\usepackage[inline]{enumitem}
\usepackage[justification=justified, skip=5pt]{caption} %

\renewcommand{\arraystretch}{0.85}

\makeatletter
\g@addto@macro\normalsize{%
  \abovedisplayskip 6pt plus 4pt minus 3pt%
  \belowdisplayskip \abovedisplayskip
  \abovedisplayshortskip 6pt plus4pt  minus3pt%
  \belowdisplayshortskip 6pt plus4pt minus3pt%
}

\makeatother

\newcommand{\ourmodel}{LLMCRS\xspace}
\newcommand{\ourmethod}{RLPF\xspace}
\newcommand{\partitle}[1]{\noindent\textbf{#1}}

\newcommand\blfootnote[1]{%
  \begingroup
  \renewcommand\thefootnote{}\footnote{#1}%
  \addtocounter{footnote}{-1}%
  \endgroup
}

\AtBeginDocument{%
  \providecommand\BibTeX{{%
    \normalfont B\kern-0.5em{\scshape i\kern-0.25em b}\kern-0.8em\TeX}}}

\setcopyright{acmcopyright}
\copyrightyear{2024}
\acmYear{2024}
\acmDOI{XXXXXXX.XXXXXXX}

\acmConference[Conference acronym 'XX]{Make sure to enter the correct
  conference title from your rights confirmation emai}{June 03--05,
  2018}{Woodstock, NY}
\acmPrice{15.00}
\acmISBN{978-1-4503-XXXX-X/18/06}

\begin{document}

\title{A Large Language Model Enhanced Conversational Recommender System}

\author{Yue Feng$^{\dagger *}$\quad Shuchang Liu$^\S$ \quad Zhenghai Xue$^\ddagger$ \quad Qingpeng Cai$^\S$ \\ Lantao Hu$^\S$ \quad Peng Jiang$^\S$ \quad Kun Gai$^\heartsuit$ \quad Fei Sun$^{\diamond \clubsuit}$}
\affiliation{
\institution{
  $^\dagger$University College London, London, UK \quad $^\S$Kuaishou Technology \\
  $^\ddagger$Nanyang Technology University, Singapore  \quad $^\heartsuit$Unaffliated\\
  $^\diamond$Institute of Computing Technology, Chinese Academy of Sciences, China}\country{}
  {$^\dagger$yue.feng.20@ucl.ac.uk} \quad{$^\ddagger$zhenghai001@e.ntu.edu.sg} \quad  
  {$^\diamond$sunfei@ict.ac.cn} \\
  {$^\S$\{liushuchang, caiqingpeng, hulantao, jiangpeng\}@kuaishou.com} \quad
  {$^\heartsuit$gai.kun@qq.com}}


\renewcommand{\authors}{Yue Feng, Shuchang Liu, Zhenghai Xue, Qingpeng Cai, Lantao Hu, Lantao Hu, Peng Jiang, Kun Gai, and Fei Sun}

\renewcommand{\shortauthors}{Yue Feng, Shuchang Liu, et al.}
\begin{abstract}
Conversational recommender systems (CRSs) aim to recommend high-quality items to users through a dialogue interface. 
It usually contains multiple sub-tasks, such as user preference elicitation, recommendation, explanation, and item information search.
To develop effective CRSs, there are some challenges:
1) how to properly manage sub-tasks;
2) how to effectively solve different sub-tasks; and 
3) how to correctly generate responses that interact with users. 
Recently, Large Language Models (LLMs) have exhibited an unprecedented ability to reason and generate, presenting a new opportunity to develop more powerful CRSs.
In this work, we propose a new LLM-based CRS, referred to as \ourmodel, to address the above challenges.
For sub-task management, we leverage the reasoning ability of LLM to effectively manage sub-task. 
For sub-task solving, we collaborate LLM with expert models of different sub-tasks to achieve the enhanced performance. 
For response generation, we utilize the generation ability of LLM as a language interface to better interact with users.
Specifically, \ourmodel divides the workflow into four stages: sub-task detection, model matching, sub-task execution, and response generation. 
\ourmodel also designs schema-based instruction, demonstration-based instruction, dynamic sub-task and model matching, and summary-based generation to instruct LLM to generate desired results in the workflow.
Finally, to adapt LLM to conversational recommendations, we also propose to fine-tune LLM with reinforcement learning from CRSs performance feedback, referred to as \ourmethod. 
Experimental results on benchmark datasets show that \ourmodel with \ourmethod outperforms the existing methods.
\blfootnote{$^*$ Work done while Yue Feng was an intern at Kuaishou Technology.}
\blfootnote{$^{\clubsuit}$ Corresponding author.}
\end{abstract}






\maketitle

\input{section/introduction}

\input{section/related_work}
\input{section/framework}

\input{section/rl}

\input{section/experiment}

\input{section/discussion}
\input{section/conclusion}


\section{Ethical Considerations}
Ethical considerations for conversational recommendation systems are important. 
As these systems interact directly with users and influence their decisions, it is essential to ensure transparency, fairness, and user privacy. Striking a balance between providing personalized recommendations and avoiding manipulation or discrimination is paramount. Careful attention must be paid to data collection and usage, as well as the potential for biases in the data. Additionally, user consent, data security, and the potential impacts on individuals and society must be thoroughly evaluated and addressed. By actively addressing ethical concerns, developers can enhance user experiences while upholding ethical principles and social responsibilities.


\bibliographystyle{ACM-Reference-Format}
\bibliography{sample-base}
\balance
\appendix

\end{document}

%% file: section/introduction.tex
\section{Introduction}\label{sec:introduction}
\begin{figure}[!t]
\centering
\includegraphics[scale=0.43]{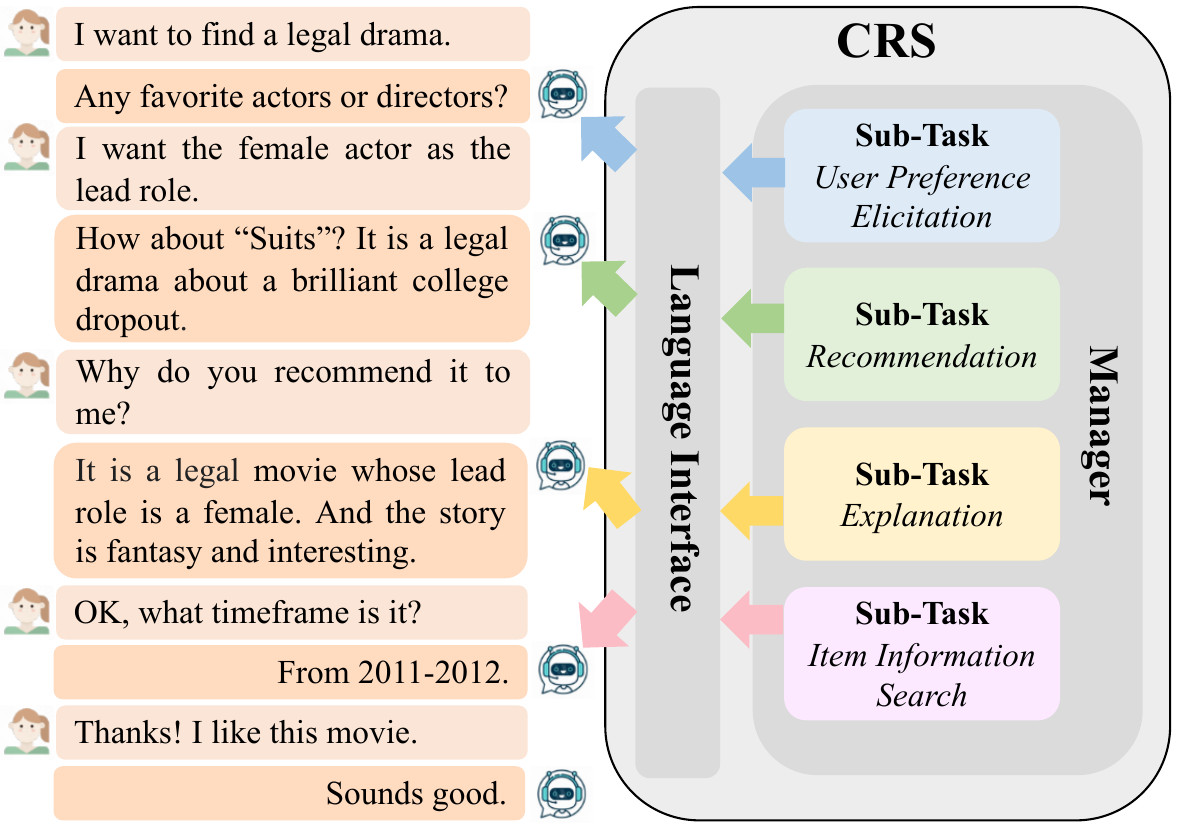}
\caption{An example of a conversational recommender system (CRS). A conversational recommendation typically consists of multiple sub-tasks. A CRS must have the ability for sub-task management and sub-task resolution. It also needs to generate responses to interact with users.}
\label{fig:example}
\end{figure}

With the advancements in conversational intelligence in recent years, conversational recommender systems (CRSs) have attracted much more attentions~\citep{chen2017survey,gao2018neural}. 
It usually consists of several sub-tasks, including user preference elicitation, recommendation, explanation, and item information search~\cite {gao2021advances}.
Figure~\ref{fig:example} shows an example of CRSs where a user resorts to the agent for movie suggestions. 
Combining the user’s preference elicited through conversational interactions, the system can offer desired recommendations easily. 
The system can also answer user's questions about recommendations such as why to recommend this item and more information about the item. 

To develop more effective CRSs, there are some challenges that need to be addressed:
1) First, the CRSs need a sub-task manager. 
Given that multiple sub-tasks are involved in conversational recommendation in a consistent dialog interface, the CRSs should effectively decide when to perform which sub-task. 
2) The CRSs need to proficiently address various sub-tasks individually. The CRSs need to design and optimize for specific sub-tasks, allowing them to achieve enhanced performance.
And 3) the CRSs need to have superior language generation ability. The CRSs should produce fluent, consistent, and meaningful responses to the users.

Recently, due to the remarkable understanding and generation capabilities demonstrated by the Large Language Models (LLMs)~\citep{brown2020language,ouyang2022training,chowdhery2022palm,zeng2022glm,zhang2022opt,touvron2023llama}, new possibilities emerge for creating more powerful CRSs. 
There are a few preliminary approaches use LLMs in CRSs, such as using the understanding ability of LLMs to rerank the results of recommendation model~\citep{wang2023rethinking} and using the generation ability of LLMs to enhance the generated responses~\citep{gao2023chat} . 
However, they cannot conduct sub-task management and they also cannot effectively address different sub-tasks, resulting in an inferior CRSs.
Inspired by some recent advances that use LLMs for task planning, external tools plug-in, and generation~\citep{schick2023toolformer,shen2023hugginggpt}, 
we propose a novel framework for CRSs, referred to as \textit{\ourmodel}, which 
1) utilizes LLMs to effectively manage sub-tasks,
2) collaborates LLMs with several expert models of different sub-tasks to improve sub-task performance; and
3) uses LLM as a language interface to generate enhanced responses to the users.
The first challenge of \ourmodel is how to define the workflow so that the LLM can simultaneously enhance sub-task management, facilitate collaboration with expert models, and optimize response generation. 
The second challenge of \ourmodel is how to effectively instruct LLM to generate specific and desired outputs in the workflow to improve the controllability of the system.
The third challenge of \ourmodel is how to adapt LLM to conversational recommendation data to improve the overall performance of the system.

Specifically, for the first challenge, the workflow of \ourmodel is divided into four stages: \textit{sub-task detection}, \textit{model matching}, \textit{sub-task execution}, and \textit{response generation}, as shown in Figure~\ref{fig:framework}, which can simultaneously enhance sub-task management, facilitate collaboration with expert models, and optimize response generation by LLM. 
For the second challenge, \ourmodel designs different mechanisms at each stage to effectively instruct LLM to generate desired results.
1) In the task detection stage, \ourmodel uses \textit{schema-based instruction} and \textit{demonstration-based instruction} to instruct LLM to understand the criteria for sub-task detection. 
2) In the model matching stage, \ourmodel uses a \textit{dynamic sub-task and model matching mechanism} to instruct LLM to select the suitable expert model for the sub-task. 
3) In the task execution stage, \ourmodel calls the selected expert model from hybrid inference endpoints to execute the sub-task. 
4) In the response generation stage, \ourmodel uses the \textit{summary-based generation} to instruct LLM to incorporate all the information of the previous stages to generate responses.
For the third challenge, we propose a new method, referred to as \textit{\ourmethod}, which utilizes reinforcement learning from CRSs performance feedback to fine-tune LLMs to adapt to conversational recommendation data to achieve better performance.
\ourmethod uses \textit{recommendation feedback} and \textit{conversation feedback} as reward signals and uses the REINFORCE~\citep{li2017deep} method to guide the learning direction. 
In order to lead to more stable updates and learning, we also employ a baseline function to reduce the variance of the estimated reward.


We conduct experiments on two benchmark datasets for conversational recommendation, which are GoRecDial and TG-ReDial datasets. The experimental results show that \ourmodel substantially outperforms existing methods on recommendation and also provides a more satisfying natural language interaction. 
The extensive analysis also reveals the significance of sub-task management, the impact of cooperating with expert models of sub-tasks, the efficacy of utilizing reinforcement learning from CRSs performance feedback, and the effectiveness of instruction mechanisms for LLM.

To sum up, our contributions are as follows:
\begin{itemize}[leftmargin=18pt, itemsep=3pt] 
\item We propose a new framework \ourmodel which effectively utilizes LLM to solve challenges in CRSs, including sub-task management, proficient sub-tasks handling, and advanced language generation.
\item \ourmodel defines a new workflow for CRSs and designs various mechanisms to efficiently instruct LLMs within this workflow.
\item \ourmodel is refined by reinforcement learning from CRSs performance feedback (\ourmethod) to get an enhanced performance.
\item The experiment results of \ourmodel achieve the best performances against baselines on recommendation and have better natural language interaction. Extensive experimental analysis also helps to better understand the advantages of our method.
\end{itemize}

%% file: section/related_work.tex
\section{Related Work}\label{sec:related}

\subsection{Conversational Recommender System}
Conversational recommender systems (CRSs) aim to provide recommendation services through conversational interactions. Two main types of CRSs have been studied: attribute-based CRSs and generation-based CRSs. 
Attribute-based CRSs interact with users through pre-defined actions~\citep{lei2020estimation}. They capture user preferences by asking queries about item attributes and generate responses using pre-defined templates~\citep{shang2023multi,zhou2020leveraging}.
They mainly focus on capturing user preferences and giving accurate recommendations within as few turns as possible. 
Compared to attribute-based CRSs, generation-based CRSs interact with users through more free-form natural language conversations~\citep{li2018towards}.
They capture the preferences from the conversation context and then generate responses with free-form~\citep{chen2019towards,zhou2020towards,zhou2020improving}. 
They mainly focus on giving accurate recommendations through a natural language interface. 
Such approaches often utilize an end-to-end framework~\citep{li2018towards,tu2022conversational,zhou2021crfr,zhou2022c2}. However, due to the intricate nature of CRSs, their effectiveness is limited.
In this study, we focus on the type of generation-based CRSs. Our objective is to decompose intricate generation-based CRSs into discrete sub-tasks, thereby enhancing performance for each individual sub-task.

\subsection{LLMs for Recommendation}
With the emergence of LLMs in natural language processing, there has been a growing interest in harnessing the power of LLMs to enhance recommendation~\citep{wu2023survey,liu2023pre}. 
The LLMs are used for two types of recommendation systems: traditional recommendation and conversational recommendation. 
Traditional recommendations primarily predict a user’s preference toward an item by analyzing user past behaviors.
For the traditional recommendation, they propose to utilize the LLMs as a feature extractor, which feeds the context of items and user past behaviors into LLMs and outputs corresponding embeddings~\citep{zhang2023recommendation,wang2023zero}. 
Some approaches directly use LLMs as recommendation systems\citep{dai2023uncovering,zhang2023chatgpt}. The input sequence usually consists of the user profile description, item description, and interaction history. The output sequence is expected to offer a recommendation item.
For the conversational recommendation, there are only some preliminary approaches. 
Some researchers propose to utilize in-context learning to rerank the results of the recommendation model~\citep{wang2023rethinking}. 
Some researchers propose to utilize prompting strategies to generate the responses~\citep{wang2023rethinking}. 
However, they ignore the decomposition problem in the CRSs. In this work, we focus on a conversational recommendation. We target using LLMs to decompose CRSs into sub-tasks. 

%% file: section/framework.tex
\section{Framework}\label{sec:framework}

\begin{figure*}
\centering
\includegraphics[scale=0.70]{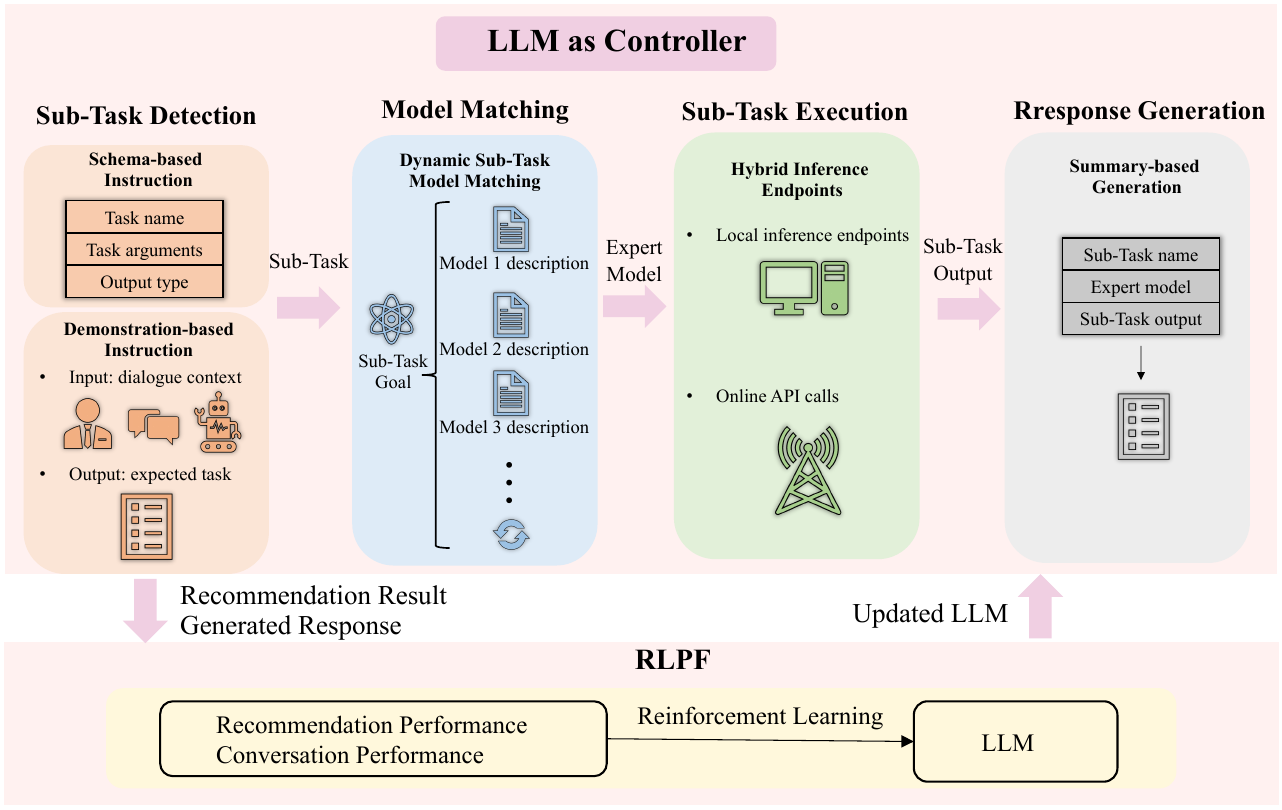}
\caption{The framework of \ourmodel. It consists of an LLM as the controller and numerous expert models as collaborative executors. The LLM ﬁrst detects the sub-task. Then it selects a suitable expert model for the sub-task. After that, the sub-task is assigned to the expert model to get the results. Finally, LLM collects the results from the expert model and generates responses to the user. The Reinforcement Learning from CRSs Performance Feedback (\ourmethod) is used to adapt the LLM to CRSs.}
\label{fig:framework}
\end{figure*}

\ourmodel is a collaborative conversational recommender system that consists of an LLM as the manager and numerous expert models as collaborative executors. The workflow of \ourmodel consists of four stages: sub-task detection, model matching, sub-task execution, and response generation, as shown in Figure~\ref{fig:framework}. Specifically, 1) An LLM first analyzes the dialogue context to detect which sub-task should be executed for the current dialogue turn based on its knowledge; 2) The LLM distributes the detected sub-task to a suitable expert model according to the model descriptions; 3) The expert model executes the assigned sub-task on the inference endpoint and returns the execution information and inference results to LLM; 4) Finally, the LLM summarizes the execution information and inference results, and generates the response to the user.

\subsection{Sub-Task Detection}
In the first stage of \ourmodel, the LLM takes the dialogue context as input and detects which sub-task should be executed for the current dialogue turn. Complex conversational recommender systems often involve several sub-tasks and different sub-tasks requiring different sub-task arguments. Schema can be used to provide a uniform template for different sub-tasks. And demonstrations can also instruct LLM to get better performance.
Therefore, to unify different tasks and improve the LLM's understanding of the criteria for sub-task detection, \ourmodel employs schema-based instruction and demonstration-based instruction in its prompt design. 
The prompt is shown in Table~\ref{tab:prompt_design}. We introduce the details in the following paragraphs.

\begin{table}
\centering
\renewcommand{\arraystretch}{1.15}
\caption{Sub-tasks schema in \ourmodel.}  
\begin{adjustbox}{width=\linewidth}
\begin{tabular}{c|c|c}
        \toprule
        \bf{Sub-Task Name} & \bf{Sub-Task Arguments} & \bf{Output type}  \\
        \midrule
        \text{User Preference Elicitation} &  \{category, dialogue context\} & Text\\
        \text{Recommendation} & \{category, dialogue context\} & Item\\
        \text{Explanation} & \{item name, dialogue context\} & Text\\
        \text{Item Detail Search} & \{item name, item attribute\} & Text\\
        \toprule
	\end{tabular}
\end{adjustbox}
\label{tab:task}
\end{table}

\partitle{Schema-based Instruction}: 
Schema can be used to provide a uniform template for different sub-tasks in the conversational recommender system. \ourmodel designs three slots in the schema for sub-task detection, which are the sub-task name, sub-task arguments, and output type:
\begin{itemize}[leftmargin=18pt, itemsep=3pt] 
\item \textit{Sub-task name}: It provides a unique identifier for sub-task detection, which is used for references to related expert models and their predicted results. The list of currently supported subtasks of \ourmodel is shown in Table~\ref{tab:task}.
\item \textit{Sub-task arguments}: It contains the list of required arguments for sub-task execution. They are resolved from either the dialogue context or the predicted results of the sub-tasks executed in the previous dialogue turns. The corresponding arguments for different sub-tasks are shown in Table~\ref{tab:task}.
\item \textit{Output type}: It indicates the type of the outputs of the sub-task. The corresponding types of outputs for different sub-tasks are shown in Table~\ref{tab:task}.
\end{itemize}

\partitle{Demonstration-based Instruction}:
By injecting several demonstrations into the prompts, \ourmodel can allow the large language model to better understand the criteria for sub-task detection. Each demonstration is a set of inputs and outputs on sub-task detection, which are the dialogue context and the expected sub-tasks to be executed. The sub-task in the demonstrations is represented by schema, which is inferred by the LLM. Demonstrations can effectively aid \ourmodel in understanding the criteria for sub-task detection and the semantic meaning of the slots in the schema. 

Benefiting from instruction tuning~\citep{longpre2023flan} and reinforcement learning from human feedback~\citep{ouyang2022training}, the large language model has the ability to follow instructions. \ourmodel provides these schema-based instructions and demonstration-based instructions to the large language model as high-level instructions for analyzing the dialogue context and detecting sub-tasks accordingly.

\begin{table}
\centering
\caption{The details of the prompt design in \ourmodel. There are some injectable slots in the prompts, such as Sub-Task List, Sub-Task Schema, Demonstrations, Dialogue Context, Sub-Task Goal, Expert Models, and Sub-Task Output. These slots are uniformly replaced with the corresponding text before being fed into the LLM.}  
\begin{tabular}{p{0.6cm}|p{7cm}}
        \toprule
        \bf{Stage} & \bf{Prompt}\\
 	\midrule
        \rotatebox{270}{Sub-Task Detection} &The AI assistant should analyze the dialogue context to detect which sub-task should be selected. The sub-tasks include: \textit{\{\{ Sub-Task List \}\}}. The selected sub-task need to represent by its corresponding sub-task schema. The sub-task schema is \textit{\{\{ Sub-Task Schema \}\}}. There are some cases for your reference: \textit{\{\{ Demonstrations \}\}}. The dialogue context is \textit{\{\{ Dialogue Context \}\}}. From this dialogue context, which sub-task should be selected? The sub-task MUST be selected from the above list and represented by the schema. \\
        \midrule
        \rotatebox{270}{Model Matching} & The AI assistant should select the most appropriate expert model to process the sub-task based on the sub-task goal and expert model description. The sub-task goal is \textit{\{\{ Sub-Task Goal \}\}}. The list of expert models is \textit{\{\{ [ID, Description] \}\}}. Please select one expert model. The expert model MUST be selected from the above list and represented by the ID.\\
        \midrule
        \rotatebox{270}{Response Generation} &With the dialogue context and the sub-task results, the AI assistant needs to generate a response to the user. The dialogue context is \textit{\{\{ Dialogue Context \}\}}. The sub-task results can be formed as: Sub-Task Name: \textit{\{\{ Sub-Task Name \}\}}, Expert Model: \textit{\{\{ ID, Description \}\}}, and Sub-Task Output: \textit{\{\{ Output \}\}}. Please generate a response to answer the user’s request.\\
        \bottomrule
	\end{tabular}
\label{tab:prompt_design}
\end{table}


\subsection{Model Matching}
After detecting the sub-task to be executed, \ourmodel next needs to match the sub-task and the expert model, that is to say, select an appropriate model from the candidate expert model set for the sub-task. To make \ourmodel have the scalability to the candidate expert model set, we propose a dynamic sub-task and model matching mechanism, which can select a model from the dynamic candidate expert model set.
This practice makes \ourmodel more open and flexible. The prompt is shown in Table~\ref{tab:prompt_design}.
We introduce the details in the following paragraphs.

\partitle{Dynamic Sub-Task and Model Matching}:
To make \ourmodel have incremental expert model access ability, we approach the matching of sub-tasks and models as matching problem between sub-task goal and model description, where sub-tasks and expert models are presented using text in the prompt. 
When new expert models are added to the system, we just simply add the new expert model descriptions to the prompt and analyze the relevance between the new model description and the sub-task goal. 
Model descriptions are often provided by the developers, which contain information on functions, architecture, supported languages, domains, licensing, and more. 
This information can help LLM to understand the functionality of the expert models. Therefore, presenting sub-task using its goal and presenting expert models using model descriptions in the prompt can support \ourmodel selecting the suitable model for the sub-task from a set of candidate expert models.

\subsection{Sub-Task Execution}
Once a sub-task is assigned to a specific expert model, the next step is to execute the sub-task, i.e., perform the expert model inference process. By taking the sub-task arguments at the sub-task detection stage as inputs, the expert model computes the sub-task outputs and then sends them back to the LLM. Besides, at this stage, to address the expert model availability problem and data security problem, the expert models should be run on hybrid inference endpoints, which are online API calls and local inference endpoints.


\subsection{Response Generation}
After sub-task execution is completed, \ourmodel enters the response generation stage. To better instruct LLM to generate the response to the user, \ourmodel needs to utilize dialogue context and all the information of the previous stages, including sub-task detection, model matching, and sub-task execution as inputs to the LLM. Therefore, we propose a summary-based generation practice. The prompt is shown in Table~\ref{tab:prompt_design}. We introduce the details in the following paragraphs.

\partitle{Summary-based Generation}:
To incorporate all the information of the previous stages, \ourmodel takes a summary as inputs of LLM. To make the summary concise, we represent the summary in a structured format, which includes three attributes as follows,
\begin{itemize}[leftmargin=15pt,itemsep=3pt, topsep=3pt] 
\item \textit{Sub-task name}: The detected sub-task for current dialogue turn.
\item \textit{Expert model}: The description of the selected expert model for the detected sub-task.
\item \textit{Sub-Task output}: The inference results of the expert model.
\end{itemize}
\ourmodel allows the LLM to generate the final responses using the summary and dialogue context as input. It can effectively incorporate all the information from the previous stages and the dialogue context, helping to provide more instructions to the LLM.

%% file: section/rl.tex
\section{Reinforcement Learning from CRSs Performance Feedback}\label{sec:rl}
While learning only from prompts is a powerful method for instructing LLM, it is not sufficient to solve real-world recommendation problems that require a deeper understanding of dialogue context and recommendation environment. One potential method to improve the understanding capabilities of LLM is to use reinforcement learning techniques to fine-tune LLM for conversational recommendation problems. In this work, we propose Reinforcement Learning from CRSs Performance Feedback (\ourmethod), which uses recommendation performance and response generation performance to guide LLM learning, resulting in improved overall performance for CRSs.

In the setup of \ourmethod, the environment is the proposed \ourmodel platform and the agent is the large language model $L$ parameterized with $\Theta$. The solutions $S$ generated by the LLM can be seen as actions that are used to solve the conversational recommendation problem. We can use the performance on that dataset as the reward signal and use reinforcement learning to fine-tune the LLM. More concretely, to find the optimal solution, we require the LLM to maximize its expected reward $\mathcal{R}$ on the training set $T_{train}$, represented by $J(\Theta)$:
\begin{gather}
J(\Theta) = \mathbb{E}_{S\thicksim L(T_{train}|\Theta)} [\mathcal{R}].
\end{gather}
The reward $\mathcal{R}$ is composed of the recommendation evaluation metric HIT~\citep{niu2012top} and the response generation evaluation metric BLEU~\citep{papineni2002bleu} with balance parameter $\lambda$. Here is a preliminary attempt to demonstrate the effectiveness of \ourmethod learning framework. In future work, we can also explore more other evaluation metrics.
\begin{gather}
\mathcal{R} = \lambda \cdot \text{HIT} + (1-\lambda) \cdot \text{BLEU}.
\end{gather}
Since the reward signal $\mathcal{R}$ is non-differentiable, we need to use a policy gradient method to iteratively update $\Theta$. In this work, we use the REINFORCE~\citep{li2017deep} method to update the parameters as follows,
\begin{gather}
\nabla_\Theta J(\Theta) = \mathbb{E}_{p(S|\Theta)} [\nabla_\Theta \text{log} P(S|\Theta)\cdot \mathcal{R}].
\end{gather}
An empirical approximation of the above quantity as follows,
\begin{gather}
\nabla_\Theta J(\Theta) \approx \frac{1}{|T_{train}|} \sum_{t \in T_{train}} \nabla_\Theta \text{log} P(S|\Theta)\cdot \mathcal{R}.
\end{gather}
The above update is an unbiased estimate for the gradient but has a very high variance. In order to reduce the variance of this estimate, following previous work~\citep{sutton2018reinforcement}, we employ a baseline function $b$, which is a moving average of the previous reward signals:
\begin{gather}
\nabla_\Theta J(\Theta) \approx \frac{1}{|T_{train}|} \sum_{t \in T_{train}} \nabla_\Theta \text{log} P(S|\Theta)\cdot (\mathcal{R}-b(S)).
\end{gather}

The RLPF approach can effectively refine the large language models to recommendation and response generation, leading to a significant improvement for  CRSs.

\begin{table}
\centering
\renewcommand{\arraystretch}{1.1}
\caption{Statistics of GoRecDial and TG-ReDial in the experiments.}
\resizebox{0.42\textwidth}{!}{
\begin{tabular}{l|cc}
        \toprule
        \textbf{Characteristics}&\textbf{GoRecDial}&\textbf{TG-ReDial}\\
 		\midrule
        \text{\#Dialogues}&  9,125 & 10,000\\
        \text{\#Utterances} &170,904& 129,392\\
        \text{Avg turn per dialogue} & 9.3& 6.5\\
        \text{Avg token per utterance} & 8.4& 19.0\\
        \text{\#Item}& 5,300 &33,834\\
		\bottomrule
	\end{tabular}
	}
\label{tab:dataset}
\end{table}

\begin{table*}
\centering
\caption{Performance of \ourmodel and baselines for recommendation on GoRecDial. Numbers in \textbf{bold} denote the best results in that metric. \ourmodel significantly improves over the best baseline (two-sided paired t-test, $p < 0.05$).}
\resizebox{0.99\textwidth}{!}{
\begin{tabular}{l|ccc|ccc|ccc}
        \toprule
        \textbf{Model}&\textbf{HIT@1}&\textbf{HIT@10}&\textbf{HIT@50} &\textbf{ MRR@1}&\textbf{MRR@10}&\textbf{MRR@50} &\textbf{NDCG@1}&\textbf{NDCG@10}&\textbf{NDCG@50} \\
 		\midrule
        \text{BERT} &0.0458   &0.2212   &0.4827 &0.0458  &0.0905   &0.1024  &0.0458    &0.1212    &0.1780 \\
        \text{GPT-2}& 0.0111   &0.1095   &0.2927 &0.0111  &0.0305  &0.0376  &0.0111   &0.0484    &0.0866 \\
        \text{ReDial} & 0.0483 &  0.2185   &0.4599 &0.0483  &0.0948  &0.1123  &0.0483   &0.1295   &0.1805\\
        \text{KBRD} & 0.0581   &0.2268   &0.4626  &0.0581   &0.1037   &0.1150   &0.0581    &0.1326   & 0.1848\\
        \text{KGSF} & 0.0592   &0.2603   &0.5374    & 0.0592   &0.1074   &0.1206  &0.0592    &0.1429    &0.2042\\
        \text{TG-ReDial} & 0.0519   &0.2853   &0.5496 &0.0519   &0.1138   &0.1267  &0.0519   &0.1539    &0.2129\\
        \midrule
        \text{\bf \ourmodel(Flan-T5)} & 0.0612  & 0.3022 & 0.5714 & 0.0612 &  0.1253 &  0.1501&0.0612 &   0.1723  &  0.2286\\
            \text{\bf \ourmodel(LLaMA)} & \textbf{0.0635}  &\textbf{0.3031}  &\textbf{0.5718}  &\textbf{0.0635}  &\textbf{0.1296}  &\textbf{0.1518}  &\textbf{0.0635}  &\textbf{0.1783} &\textbf{0.2290}\\
		\bottomrule
	\end{tabular}
	}
\label{tab:result_recommendation_g}
\end{table*}

\begin{table*}
\centering
\caption{Performance of \ourmodel and baselines for recommendation on TG-ReDial. Numbers in \textbf{bold} denote the best results in that metric. \ourmodel significantly improves over the best baseline (two-sided paired t-test, $p < 0.05$).}
\resizebox{0.99\textwidth}{!}{
\begin{tabular}{l|ccc|ccc|ccc}
        \toprule
        \textbf{Model}&\textbf{HIT@1}&\textbf{HIT@10}&\textbf{HIT@50} &\textbf{ MRR@1}&\textbf{MRR@10}&\textbf{MRR@50} &\textbf{NDCG@1}&\textbf{NDCG@10}&\textbf{NDCG@50} \\
 		\midrule
        \text{BERT} & 0.0072	&0.0049	&0.0281	&0.0072	&0.0106	&0.0124	&0.0049	&0.0147	&0.0239\\
        \text{GPT-2}& 0.0021 & 0.0192 &0.0421 & 0.0021 & 0.0051 & 0.0082 & 0.0021 & 0.0102 & 0.0187\\
        \text{ReDial} &0.0028 &0.0249 &0.0533 &0.0028 &0.0073 &0.0104 &0.0028 &0.0112 &0.0203\\
        \text{KBRD} &0.0040&	0.0254	&0.0588	&0.0040&	0.0089&	0.0103	&0.0040&	0.0127&	0.0198 \\
        \text{KGSF} &0.0053&	0.0285	&0.0771	&0.0053&	0.0114	&0.0135	&0.0053&	0.0154&	0.0259\\
        \text{TG-ReDial} & 0.0079	&0.0251	&0.0524	&0.0079&	0.0122	&0.0134	&0.0079	&0.0152	&0.0211\\
        \midrule
        \text{\bf \ourmodel(Flan-T5)} & 0.0084 & 0.0302& \textbf{0.0792} & 0.0084 & 0.0128 & 0.0138 & 0.0084 & 0.0159 & 0.0261\\
        \text{\bf \ourmodel(LLaMA)} & \textbf{0.0086} & \textbf{0.0308} & 0.0791 & \textbf{0.0086} & \textbf{0.0130} & \textbf{0.0139} & \textbf{0.0086} & \textbf{0.0162} & \textbf{0.0263}\\
		\bottomrule
	\end{tabular}
	}
\label{tab:result_recommendation_t}
\end{table*}

\begin{table}
\centering
\caption{Performance of \ourmodel and baselines for conversation on GoRecDial. Numbers in \textbf{bold} denote the best results in that metric. \ourmodel significantly improves over the best baseline on Distinct (two-sided paired t-test, $p < 0.05$).}
\begin{adjustbox}{width=\linewidth}
\begin{tabular}{l|cccc}
        \toprule
        \textbf{Model}&\textbf{BLEU-1}&\textbf{BLEU-2}&\textbf{Distinct-1}&\textbf{Distinct-2} \\
 		\midrule
        \text{GPT-2} &0.0120 & 0.0020& 0.0219  &0.1542 \\
        \text{ReDial} &0.0421& 0.0035 & 0.0044  & 0.0124\\
        \text{KBRD} &\textbf{0.0782}&\textbf{0.0068} & 0.0081 & 0.0235  \\
        \text{KGSF} &0.0467& 0.0037  & 0.0078  & 0.0124\\
        \text{TG-ReDial} & 0.0226 &0.0046 & 0.0120  &0.0960\\
        \midrule
        \text{\bf \ourmodel(Flan-T5)} & 0.0528 & 0.0050 & \textbf{0.1659} & \textbf{0.2944}\\
        \text{\bf \ourmodel(LLaMA)} & 0.0601   &0.0057 
 &0.1590  &0.2903\\
		\bottomrule
	\end{tabular}
\end{adjustbox}
\label{tab:result_conversation_g}
\end{table}

%% file: section/experiment.tex
\section{Experiments}\label{sec:experiment}
\subsection{Dataset}
We conduct the experiments using the benchmark datasets on conversational recommendation, namely GoRecDial~\citep{kang2019recommendation} and TG-ReDial~\citep{zhou2020towards}. The introduction of the datasets is as follows.

\partitle{GoRecDial}: It is a conversational recommendation dataset released by Kang et al.~\citep{kang2019recommendation}. This dataset was constructed using ParlAI~\citep{miller2017parlai} to interface with Amazon Mechanical Turk (AMT)\footnote{\href{https://www.mturk.com/}{https://www.mturk.com/}}. 
To reflect the movie preferences of real users, this dataset built the pool of recommendation movies using the MovieLens dataset\footnote{\href{https://grouplens.org/datasets/movielens/}{https://grouplens.org/datasets/movielens/}}, comprising 27M ratings applied to 58K movies by 280K real users. To obtain the movie information, they obtained the descriptive text for each movie from Wikipedia\footnote{\href{https://dumps.wikimedia.org/}{https://dumps.wikimedia.org/}} and extracted entity-level features (e.g., directors, actors, year) using the MovieWiki dataset~\citep{miller2016key}. The statistics of GoRecDial are presented in Table~\ref{tab:dataset}. 

\partitle{TG-ReDial}: It is a conversational recommendation dataset released by Zhou et al.~\citep{zhou2020towards}. 
The conversation was created in a semi-automatic way by involving reasonable and controllable human annotation efforts~\citep{wu2017sequential}. The movie watching records was collected from real users on Douban\footnote{\href{https://www.douban.com/}{https://www.douban.com/}} website. 
The dataset contains 1,482 users and 202.7 watching records for each user on average. The movie information was extracted from movie tags on Douban (e.g. genre, director, and starring). 
The statistics of TG-ReDial are presented in Table~\ref{tab:dataset}. 

\subsection{Baselines}
Following~\citep{zhou2020towards,zhou2020improving,chen2019towards}, we evaluate the superiority of our method by considering the following representative baselines:





\partitle{BERT}~\citep{kenton2019bert}: It is a bidirectional language model pre-trained via the masked language modeling task on a large-scale general corpus. We utilize the representation of the [CLS] token for recommendation. 



\partitle{GPT-2}~\citep{radford2019language}:  It is an autoregressive language model pre-trained via the language modeling task on large-scale general corpora. We take the generated text of language model for response and use the representation of the last token for recommendation.

\partitle{ReDial}~\citep{li2018towards}:  It includes a conversation module based on sequence-to-sequence learning and a recommendation module based on a denoising auto-encoder.

\partitle{KBRD}~\citep{chen2019towards}: It utilizes a knowledge graph to enhance the semantics of contextual items or entities for recommendation. The dialog generation module is based on the Transformer~\citep{vaswani2017attention} architecture.


\partitle{KGSF}~\citep{zhou2020improving}: It incorporates both word-oriented and entity-oriented knowledge graphs to enhance the performance of conversational recommendation.

\partitle{TG-ReDial}~\citep{zhou2020towards}: It adopts BERT to encode the historical utterances to model the dialogue topic, and leverages graph neural networks for topic-guided item recommendation and response generation.


\subsection{Evaluation Measures}
Following existing work~\citep{zhou2020towards,zhou2020improving,chen2019towards}, we adopt different metrics to evaluate the performance of recommendation and conversation.
For the recommendation, we develop ranking-based metrics for measuring the ranking performance of the generated recommendation lists, which include HIT@k~\citep{niu2012top}, MRR@k~\citep{voorhees2003overview}, and NDCG@k~\citep{jarvelin2017ir} (k = 1, 10, and 50).
For the conversation, we use both relevance-based and diversity-based evaluation metrics to measure the performance of generated responses. The relevance-based metrics include BLEU~\citep{papineni2002bleu} which measure the similarity between ground truth and generated responses from the perspective of probability. The diversity-based metrics are Distinct~\citep{li2016diversity}, measuring the number of distinct in the generated responses.

\begin{table}
\centering
\caption{Performance of \ourmodel and baselines for conversation on TG-ReDial. Numbers in \textbf{bold} denote the best results in that metric. \ourmodel significantly improves over the best baseline on Distinct (two-sided paired t-test, $p < 0.05$).}
\begin{adjustbox}{width=\linewidth}
\begin{tabular}{l|cccc}
        \toprule
        \textbf{Model}&\textbf{BLEU-1}&\textbf{BLEU-2}&\textbf{Distinct-1}&\textbf{Distinct-2} \\
 		\midrule
        \text{GPT-2} &0.0858 &0.0119& 2.3500	&4.6200\\
        \text{ReDial} & 0.0570 &0.0044&  0.0041 & 0.0070 \\
        \text{KBRD} & 0.2670 &0.0458& 0.4690 & 1.5000	\\
        \text{KGSF} & \textbf{0.3830}& \textbf{0.1150} & 0.3400 & 0.9100\\
        \text{TG-ReDial} &0.1250 &	0.0204& 0.8810	&1.7500 \\
        \midrule
        \text{\bf \ourmodel(Flan-T5)} & 0.3011 & 0.1021 & \textbf{2.4231} & \textbf{4.8332}\\
        \text{\bf \ourmodel(LLaMA)} & 0.3123 & 0.1088 & 2.4128 & 4.8023\\
		\bottomrule
	\end{tabular}
\end{adjustbox}
\label{tab:result_conversation_t}
\end{table}

\subsection{Implementation Details}
For fair comparisons, we implement all the baselines and \ourmodel by the open-source toolkit CRSLab\footnote{\href{https://github.com/RUCAIBox/CRSLab}{https://github.com/RUCAIBox/CRSLab}}~\citep{zhou2021crslab}. The hyper-parameter settings of the baselines follow the default settings on CRSLab, which reaches the best performances.
The data preprocessing is also consistent with that of CRSLab, ensuring a fair and equitable comparison.
The expert models we use in \ourmodel are the state-of-the-art methods for each task. Specifically, for the user preference elicitation task, we use KBRD~\citep{chen2019towards} and KGSF~\citep{zhou2020improving} methods; for the recommendation task, we use TG-ReDial~\citep{zhou2020towards} method; for the explanation task, we use KBRD~\citep{chen2019towards} and KGSF~\citep{zhou2020improving} methods; for the item detail search task, because the results are searched directly from the database, we do not use any expert models. 

In addition, we employ our system using two open-source large language models, which are Flan-T5-\textit{Large}~\citep{chung2022scaling} and LLaMA-\textit{7b}~\citep{touvron2023llama}.

\partitle{Flan-T5}: It is a series of language models developed by Google. It is fine-tuned using instruction-tuning, which allows them to have good performance on a variety of tasks. In our work, we use Flan-T5-\textit{Large}, which has 770 million parameters.

\partitle{LLaMA}: It is a lightweight, open-source language model developed by researchers at Meta. It is designed to be efficient and performant. In our work, we use LLaMA-\textit{7b}, which has 7-billion parameters.

Moreover, we use the Adam~\citep{kingma2014adam} optimizer to update the parameters. The learning rate and the weight decay rate are set to be 5e-5 and 0.01, respectively. The batch size is 1. The max source sequence length and the max target sequence length are 512 and 100, respectively. We train the model on eight NVIDIA GeForce RTX 2080-12GB GPUs and the training time is around 36h.

%% file: section/discussion.tex
\section{Experimental Results and Discussion}\label{sec:discussion}
\subsection{Overall Performance}
Tables~\ref{tab:result_recommendation_g}, ~\ref{tab:result_recommendation_t}, ~\ref{tab:result_conversation_g}, and ~\ref{tab:result_conversation_t} show the performance of \ourmodel as well as the baselines of recommendation and conversation on GoRecDial and TG-ReDial datasets respectively. It is shown that \ourmodel achieves state-of-the-art performance in the recommendation which is the most important performance for CRSs. All improvements observed compared to the baselines are statistically significant according to two sided paired t-test (p $<$ 0.05). 
And \ourmodel can also provide a more satisfying language interface compared to the state-of-the-arts. The performance of \ourmodel on relevance-based conversation evaluation metric is similar to the baselines. Most notably, \ourmodel can significantly improve the performance of the diversity-based evaluation metric on the conversation. It indicates that \ourmodel can keep the consistency with the ground truth dialogue and also generate more diversity and informative responses to the user, resulting in the improved interaction between the user and the system. 
We conjecture that the overall good performance of \ourmodel is due to it containing the LLM, which has exhibited exceptional ability in task planning, tool interaction, language understanding, and language generation. 
The following analysis provides a better understanding of our model’s strengths.

\begin{figure*}[!t]
\centering
\includegraphics[scale=0.56]{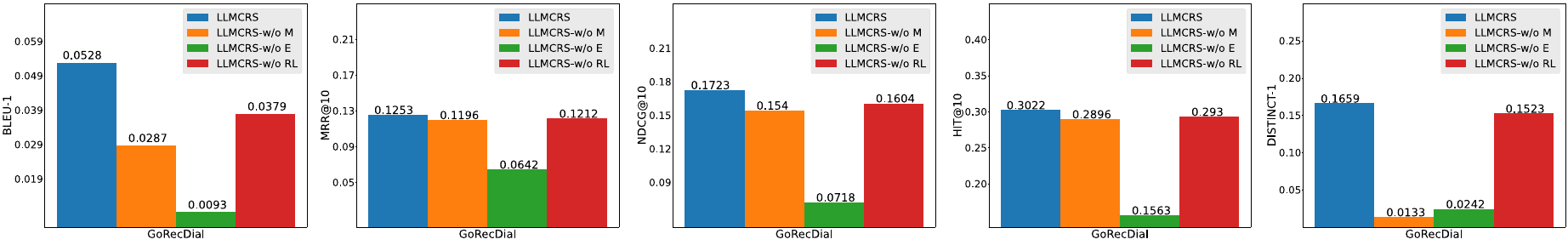}
\caption{Ablation study of \ourmodel with respect to the sub-task management,  expert models, and RLPF on GoRecDial dataset.}
\label{fig:ablation_g}
\end{figure*}

\begin{figure*}[!t]
\centering
\includegraphics[scale=0.57]{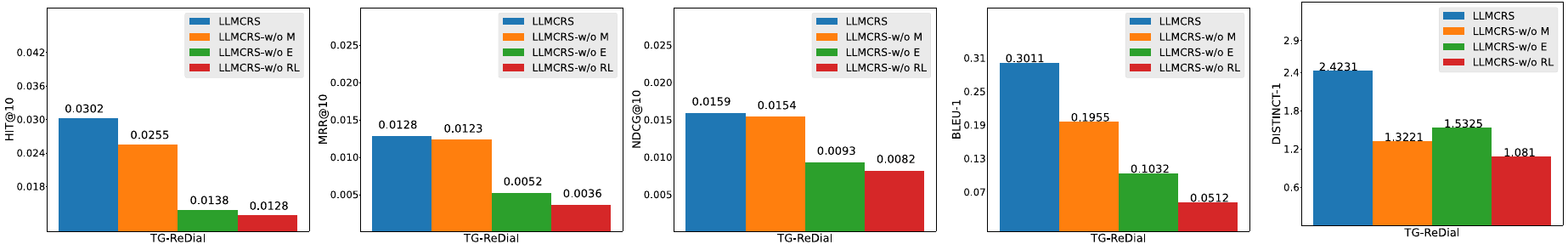}
\caption{Ablation study of \ourmodel with respect to the sub-task management, expert models, and RLPF on TG-ReDial dataset.}
\label{fig:ablation_t}
\end{figure*}

\subsection{Ablation Study}
We conduct an ablation study on \ourmodel to quantify the effects of three factors: the usage of sub-task management, the cooperation with the expert models, and the usage of the reinforcement learning from CRSs performance feedback. The results indicate that all above factors of \ourmodel are indispensable for conversational recommendation.

\partitle{Effect of the usage of sub-task management}

To investigate the effectiveness of sub-task management, we compare \ourmodel with \ourmodel-w/o M which eliminates the sub-task management in the system. 
\ourmodel-w/o M only interacts with one expert model TG-ReDial and generates responses based on the outputs of TG-ReDial, no matter which sub-task the current dialogue should solve. Figure~\ref{fig:ablation_g} and ~\ref{fig:ablation_t} show the results of \ourmodel-w/o M on GoRecDial and TG-ReDial datasets in terms of HIT@10, MRR@10, NDCG@10, BLEU-1, and Distinct-1. From the results, we can observe that without sub-task management, the performances of recommendation and conversation deteriorate considerably. It indicates that sub-task management, which effectively detects when to do which sub-task, can better understand the process of conversation and improve the overall performance of CRSs.

\partitle{Effect of Cooperation with Expert Models}

To validate the effectiveness of incorporating with the expert models, we employ each sub-task with the LLM instead of using expert models. For user preference elicitation, explanation, and item information search, we directly take the dialogue context as input of LLM and then use the responses generated by the LLM as results. For recommendation, due to the candidate item set being huge, it is hard to take all of it as input of LLM. Therefore, we first calculate the concise similarity of the item and dialogue by BERT~\citep{kenton2019bert} and select the top 50 relevant items based on the similarity score. Then, LLM takes the small set of candidates, similarity scores, and dialogue context as input to predict the recommendation.
Figure~\ref{fig:ablation_g} and ~\ref{fig:ablation_t} show the results of \ourmodel-w/o E on GoRecDial and TG-ReDial datasets. We can see that the performance of the \ourmodel-w/o E in terms of HIT@10, MRR@10, NDCG@10, BLEU-1, and Distinct-1 decreases significantly compared with \ourmodel. It indicates that cooperation with expert models can help \ourmodel to have a bridge between LLM and the task-speciﬁc models, allowing for effective knowledge transfer and improved performance. In addition, the plug-in expert model mechanism also supports dynamically adding new expert models in the system, which enhances the scalability and flexibility of CRSs.

\partitle{Effect of \ourmethod}

To analyze the effectiveness of reinforcement learning from CRSs performance feedback (\ourmethod), we compare the performance of \ourmodel with \ourmodel-w/o RL which eliminates the reinforcement learning mechanism in the system. Figure~\ref{fig:ablation_g} and ~\ref{fig:ablation_t} show the results of \ourmodel-w/o RL and \ourmodel on GoRecDial and TG-ReDial datasets in terms of HIT@10, MRR@10, NDCG@10, BLEU-1, and Distinct-1. One can observe that without \ourmethod, the performances deteriorate considerably. The \ourmethod mechanism effectively reﬁnes the LLM’s recommendation ability and response generation strategy, resulting in an enhanced and more adaptive CRS. We think that it is due to only relying on the input text for learning is insufﬁcient for LLM when solving the conversational recommendation problem. 
Performance feedback can offer valuable supplementary information that steers the learning trajectory of LLMs, enabling them to furnish more precise recommendations and generate more fitting responses. 
In addition, the absence of \ourmethod results in a more pronounced decline in the performance of TG-ReDial compared to ReDial.
We speculate that this phenomenon arises from the fact that conversations in TG-Redial are structured using predefined topic threads. Consequently, CRSs' ability to adapt to specific topics becomes crucial for achieving improved performance.
This phenomenon further demonstrates the superior adaptability of \ourmethod.

\begin{table}
\centering
\caption{Effectiveness of the mechanisms to instruct LLM in \ourmodel. Numbers in bold denote the best results.}
\begin{adjustbox}{width=\linewidth}
\begin{tabular}{p{2.6cm}|ccccc}\toprule\textbf{Model}&\textbf{HIT@10}&\textbf{MRR@10}&\textbf{NDCG@10}&\textbf{BLEU-1}&\textbf{Distinct-1} \\
 		\midrule
        \text{\bf \ourmodel(Flan-T5)} & \textbf{0.3022} & \textbf{0.1253} & \textbf{0.1723}& \textbf{0.0528} & \textbf{0.1659} \\
        $\quad$ -w/o SI & 0.2963 & 0.1228 & 0.1674& 0.0412 & 0.1588 \\
        $\quad$ -w/o DI & 0.3003 & 0.1236 & 0.1692& 0.0474 & 0.1602 \\
        $\quad$ -w/o SG & 0.2992 & 0.1235 & 0.1691& 0.0460 & 0.1591\\
		\bottomrule
	\end{tabular}
\end{adjustbox}
\label{tab:prompt}
\end{table}

\subsection{Mechanisms to Instruct LLM}
The mechanisms to instruct LLM include schema-based instruction, demonstration-based instruction, dynamic sub-task and model matching, and summary-based generation. 
Due to the limited number of  expert models we can access, we do not analyze the dynamic sub-task and model matching mechanism.
Our analysis mainly focuses on the remaining three mechanisms.
We compare \ourmodel with \ourmodel-w/o SI which removes the task schema description in the prompt, \ourmodel-w/o DI which eliminates the demonstrations in the prompt,
and \ourmodel-w/o SG which directly generates responses from the execution results of the expert models instead of the structured summary text.
Table~\ref{tab:prompt} shows the results of these methods on GoRecDial in terms of HIT@10, MRR@10, NDCG@10, BLEU-1, and Distinct-1. 
From the results, we can observe that without any mechanism in the prompts, the performances of recommendation and conversation deteriorate considerably.
We conjecture that it is due to: 
1) the schema-based instruction can unify different tasks in the same format to better guide LLM to detect different tasks and also facilitate knowledge sharing among tasks;
2) the demonstration-based instruction can condition on the in-distribution conversational recommendation demonstration to make the conversational recommendation task closer to an LLM, and demonstrations can also keep the format of the input-label pairs;
and 3) the summary-based generation can integrate all the information from the previous stages and provide global instruction to LLM for improved response generation.

\begin{figure}[!t]
\centering
\includegraphics[scale=0.5]{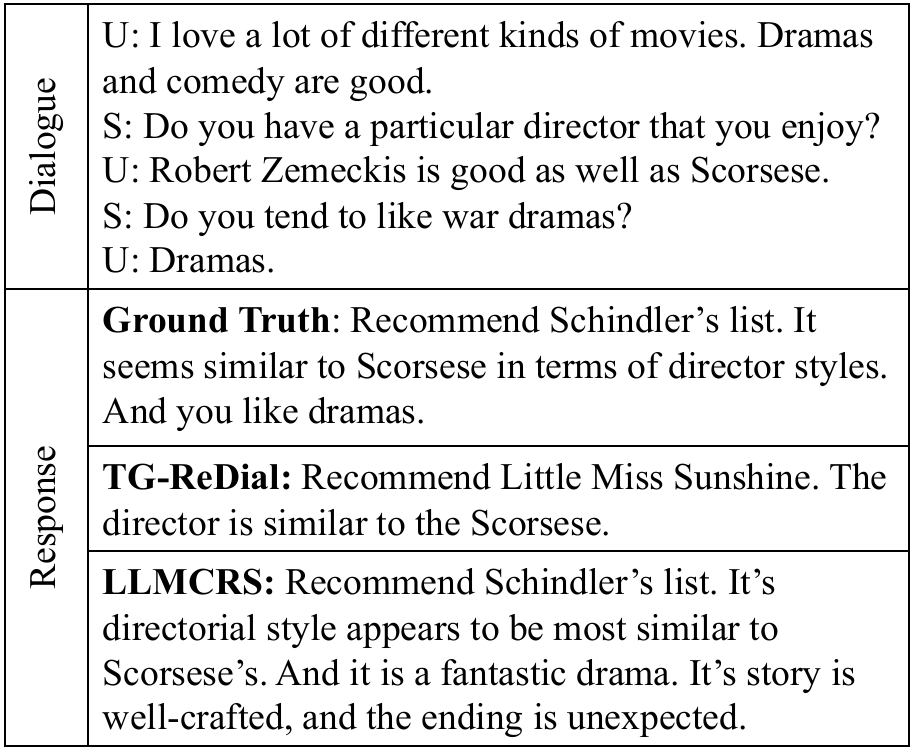}
\caption{Case study of \ourmodel}
\label{fig:case}
\end{figure}

\subsection{Case Study}
We make the qualitative analysis on the results of \ourmodel and the baselines TG-ReDial on GoRecDial datasets. We find that \ourmodel can provide accurate recommendations and generate more informative responses. For example in Figure~\ref{fig:case}, \ourmodel can coordinate the results of expert models in the previous turns on user preference elicitation task, which are "Robert Zemeckis / Scorsese" and "Dramas" to recommend "Schindler's list" to the user. \ourmodel can also leverage the language understanding and generation ability of LLM to generate more informative and interpretative responses, such as " It is a fantastic drama", "It’s story is well-crafted, and the ending is unexpected". In contrast, TG-ReDial makes a wrong recommendation. 
One potential explanation is that TG-ReDial lacks the ability to manage sub-tasks and offer solutions tailored to specific sub-tasks. Therefore, it can not accurately extract user preferences and provide accurate recommendations.
Besides, the conversation ability of TG-ReDial is inferior to the LLM, which cannot generate more informative and natural responses.

%% file: section/conclusion.tex
\section{Conclusion}\label{sec:conclusion}

We have proposed a new framework for conversational recommender systems, referred to as \ourmodel. It uses LLM to better manage sub-tasks, effectively cooperate with expert models, and generate improved responses.
The workflow of \ourmodel includes sub-task detection, model matching, sub-task execution, and response generation. At each stage, instruction learning and context learning are used to instruct LLM to perform accuratly.
We also uses reinforcement learning from CRSs performance feedback to refine LLM to provide more accurate recommendation and generate more suitable responses.
\ourmodel is a controllable and adaptable system for conversational recommendations.
Experimental results show that \ourmodel significantly outperforms the state-of-the-art methods in CRSs on the benchmark datasets of GoRecDial and TG-ReDial.
Finally, we also note that the recent rapid development of LLMs has brought a huge impact on academia and industry. We also expect the design of our model can inspire the whole community and
pave a new way for LLMs towards the recommendation.